# The Projective Wave Theory of Consciousness

Robert Worden

Theoretical Neurobiology Group, University College London, London, United Kingdom

rpworden@me.com

Draft d0.9, May 2024

Abstract:

Neural theories of consciousness face three difficulties: (1) The selection problem: how are those neurons which create consciousness selected, from all the other neurons which do not? (2) the precision problem: how do neurons hold a detailed internal model of 3D space, as the origin of our spatial conscious experience? and (3) the decoding problem: how are the many distorted neural representations of space in the brain decoded, to give our largely undistorted conscious experience of space?

These problems can all be addressed if the brain's internal model of local 3D space is held not in neurons, but in a wave excitation (holding a projective transform of Euclidean space), and if the wave is the source of spatial consciousness. Such a wave has not yet been detected in the brain, but there are good reasons why it has not been detected; and there is indirect evidence for a wave, in the mammalian thalamus, and in the central body of the insect brain.

The resulting theory of consciousness is a wave-based development of the Projective Consciousness model (PCM), called the **Projective Wave** theory of consciousness. It gives good agreement with the spatial form of our conscious experience. It has a positive Bayesian balance between the complexity of its assumptions and the data it accounts for; this gives a basis to believe it. It is falsifiable (if no wave can be found in the brain), and it gives an account of the evolution of consciousness. It implies that insects are conscious, but current man-made computers not. Possible further investigations of the theory are described.

**Keywords**: Consciousness; Bayesian assessment of theories; Bayesian Balance; spatial form of consciousness; imprecision of neural representation; wave representation of space; thalamus; insect central body; Projective Consciousness Model; Fourier transform; Bose-Einstein Condensate.

# 1. Introduction

This paper describes a theory of consciousness which agrees well with data about our conscious experience, and overcomes some problems in current neural theories of consciousness.

A scientific theory is no more than a hypothesis – and probably, an incorrect hypothesis – unless it agrees with experimental data. How much data must a theory agree with, before we can start to believe it? The answer depends on the complexity of its assumptions – the more complex the assumptions, the more data it needs to agree with. This can be quantified in the Bayesian philosophy of Science [Sprenger & Hartmann 2012]. There needs to be a positive **Bayesian Balance** between the complexity of the assumptions (a debit, which can be measured in bits) and the data the theory agrees with (a credit, also measured in bits). If the balance is positive, we may start to believe the theory.

In this regard, theories of consciousness have a difficulty. Because phenomenal consciousness is a private phenomenon, there is not much public, shareable, data to compare with any theory.

For instance, some theories only attempt to account for the existence of consciousness (David Chalmers' [1996] 'Hard Problem'). Whatever might be said about this problem, the mere existence of consciousness is not a rich source of experimental data; on its own, it offers little hope of ever confirming a theory of consciousness.

Phenomenal consciousness ('what it is like')[Nagel 1974], closely correlates with access consciousness (the ability to report information, and act on it) [Block 1978]. So as described by [Chalmers 1996] access consciousness is a source of data to test theories of phenomenal consciousness.

There are several kinds of data to test theories. This paper proposes that of all the kinds of data about consciousness, only one is a rich source of data (giving as many as thousands of bits) to test theories. That source is the spatial, geometric form of consciousness. All other sources of data have small information content by comparison, and give insufficient data to test any theory.

Therefore the only theories of consciousness which can have a positive Bayesian balance are those which predict the spatial, geometric form of consciousness – a form which is familiar to us from every moment of our waking lives. Of the many extant theories of consciousness, very few predict the spatial nature of consciousness.

Amongst neural theories of consciousness, those that do predict its spatial form include the Projective Consciousness Model (PCM) [Rudrauf et al 2017], Arnold Trehub's [1989] theory, and Tononi's [2012] IIT, as applied to spatial consciousness [Haun & Tononi 2019]. All of these are capable of building their Bayesian balance by agreeing with the data of spatial consciousness. But they have problems on the assumption debit side of their balance. These problems are of three kinds:

1. **The Selection Problem**: Out of all neural activity in the brain (most of which is not required for consciousness [Merker 2007]), what is the principle which picks out just those neurons whose activity causes consciousness? Principles have been proposed, but they have little physical basis and look like arbitrary assumptions.
2. **The Precision Problem**: Our consciousness is like a fairly precise 3-D model of the space around us; precise and detailed at least in some regions of space. If it is caused by neural activity, does that activity represent space with sufficient precision and fast response times to give consciousness as we experience it?
3. **The Decoding Problem**: The known neural representations of space in the brain are highly distorted and mostly two-dimensional. How is this neural activity decoded to remove the distortions, giving our largely undistorted conscious experience of 3-D space?

These are serious problems. This paper proposes a theory of consciousness (based on a new wave-based theory of spatial cognition) which addresses all three problems.

Spatial cognition – understanding local 3-D space, in order to control every physical movement – is the primary function of an animal brain, needed for survival at every moment of the day. Most of the brain is devoted to it. It is something of an embarrassment for neuroscience that even today, there are no working neural computational models of 3-D spatial cognition. Building working neural models of spatial cognition is a research problem widely avoided.

This may be because it is computationally a hard problem. One reason why it is hard is point (2) above. If locations in 3-D space are represented by stochastic neural firing rates, then the expected error rates (in sub-second timescales) are very high. A neuron does not give enough pulses in 1/5 second to represent space with high precision – as we know animals do, and as our own consciousness tells us we do.

This implies that we need a good computational model of 3-D spatial cognition in the brain, before we can tackle the problem of spatial consciousness.

As an alternative to storing 3-D positions as firing rates, positions of objects in local 3-space may be stored in a wave excitation in the brain, which couples to neurons. Different waves in the excitation have different wave vectors, representing different object positions in a direct and straightforward manner - much like a hologram. Wave storage can give high spatial precision and fast response times, overcoming the precision and speed problems of neural models. There is indirect evidence for such a wave in the brain, in the mammalian thalamus, and in the central body of the insect brain.

If spatial cognition works in this way, then it opens the way to a new theory of consciousness. In this theory, consciousness is caused by the wave, and is not caused by neural firing. The spatial form of consciousness is a Fourier transform of the wave amplitude (reading the hologram). This agrees well with the fact that our conscious experience of space is geometrically very like the real space around us. That is the most information-rich empirical fact about consciousness, giving the theory a strong positive Bayesian balance.

The theory overcomes the major problems of neural theories:

1. **The Selection Problem**: the theory has a very simple selection law: consciousness arises from the wave in the brain, not from any neural activity.
2. **The Precision Problem**: wave storage of positions can give high spatial precision and fast response times.
3. **The Decoding problem**: consciousness arises from the wave by a Fourier transform – a very simple decoding, familiar in many branches of science. The wave is not highly distorted.

From these points, the theory has a positive Bayesian balance, unlike any other theory of consciousness.

A wave excitation cannot directly represent locations very far from the animal; to do so it must represent some transform of Euclidean 3-D space. A nearly projective transform has the great benefit of preserving straight lines, which are used for computing motion. This theory is a wave version of the Projective Consciousness Model, and it is called the **Projective Wave Model** of consciousness.

The wave excitation has not yet been detected in the brain. This may be because the wave has evolved to have very low intensity (to save energy), and we do not yet know how to look for it. I give reasons why it is probably not an electromagnetic wave, but it may be a coherent quantum excitation. The limit of a long-lived quantum excitation would be a wave excitation of a **Bose-Einstein Condensate** (BEC). BECs (such as superfluids) can hold information for indefinite periods, so are very well suited to act as a short-term spatial memory. This possibility is an addition to the main theory, and it has significant theoretical benefits.

## 2. Bayesian Assessment of Theories of Consciousness

Since [Crick & Koch 1990; Crick1994], much has been written about phenomenal consciousness, from both philosophical and scientific viewpoints. Many theories of consciousness have been proposed. How can we assess the many competing proposals?

The proper process to evaluate a scientific hypothesis is to calculate the probability that it is correct within its domain of applicability, as a Bayesian posterior probability [Sprenger & Hartmann 2012]. Any hypothesis starts with some small prior probability of being correct, and that prior probability is modified by Bayes' theorem in the light of experimental data, to give a posterior probability. Only if the posterior probability large can we believe the hypothesis, and call it a theory.

This analysis has not been formally applied to all scientific hypotheses, but informally it is how all candidate theories are evaluated – by their fit to the data. Some established theories of physics and chemistry agree with an enormous amount of data, using only simple hypotheses; so their Bayesian posterior probability is very near 1.0, so they are taken to be correct within their domains of application. All theories and data are approximations; and the truth of a theory is not altered by new data, outside the domain where its approximations apply. So, for instance, Newtonian dynamics has not been rendered untrue by special relativity or quantum mechanics; they work in different domains.

From this Bayesian viewpoint, where do theories of consciousness stand? They are all hypotheses rather than theories. They can be assessed by a Bayesian Balance between two terms:

- Hypothesis complexity: How much information (e.g. in bits) is needed to state the hypothesis.
- Agreement with data: how significant is the agreement between the hypothesis and empirical data (which can also be measured in bits)

The balance is like an equation or inequality, where each side is approximately the log of a probability, measured in bits. Hypothesis complexity is the debit side, and agreement with the data is the credit side. Only when a hypothesis has a positive balance of credit over debit can our level of belief in it come close to 1.0.

The Bayesian Balance of theories is illustrated in figure 1:

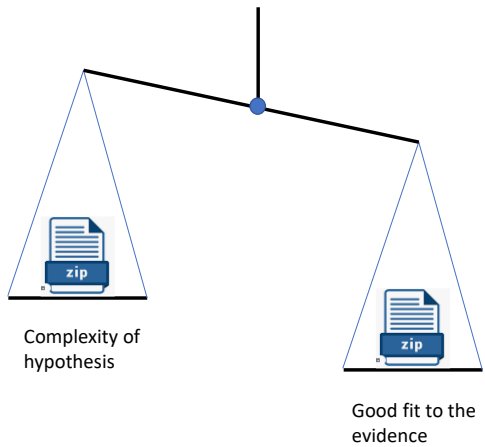

*Figure 1: The Bayesian Balance for a successful theory. The information content of the evidence fitted by a theory, as measured by the zip file test, must be greater than the information content of the hypothesis.*

The two sides of the Bayesian balance can be informally measured by a 'Zip file test': collect a statement of the hypothesis, and a statement of the evidence it agrees with, in separate text files, and compress both files using a zip compression tool. The sizes of the resulting zip files (measured in bits) are a measure of the information content of each side – the Bayesian balance.

For theories of consciousness, there are several contributions to each side of the balance. On the hypothesis side, there are three main contributions:

1. Neural/cognitive basis
2. Selection Law
3. Decoding law

**Neural/Cognitive basis**: If consciousness is assumed to be caused by events in the brain, then any theory of consciousness must be built on hypotheses about events in the brain. To the extent that the hypothesis about the brain is not confirmed by empirical data, its assumptions count as part of the assumptions of the theory of consciousness – on its debit side. If, however, the assumptions about mechanisms in the brain are independently supported by evidence, to that extent they are not treated as debit in the Bayesian balance of the theory of consciousness. However, those data, supporting a cognitive theory of the brain, cannot be taken on the credit side for the theory of consciousness; they can only reduce the debit.

**Selection Law**: Chalmers [1995,1996] showed that because conscious experience is different in kind from the physical events in brains which support cognition, if consciousness is caused by events in the brain there must be a **bridging law** between physical events and consciousness. The bridging law has two parts: a selection law and a decoding law. The selection law defines what physical events in the brain lead to consciousness (and what events do not); the decoding law defines how the properties of those physical events relate to the properties of consciousness.

This may be seen through a physical analogy. Heat is different in kind from mechanical motion; to relate the two, there needs to be a bridging law. The bridging law states that the quantity of heat is equal to the quantity of molecular kinetic energy. In this case, the selection part of the bridging law says that heat is caused by molecular kinetic energy (and not, for instance, by gravitational potential energy). The decoding law states that a property of heat - its numerical quantity - varies proportionally to the quantity of molecular kinetic energy.

The bridging law for heat is not just a verbal identity that 'heat is molecular motion'. If it was just that, there would be almost nothing to test, and no way to confirm it as a hypothesis. It is the quantitative relation that made the experiments of Joule and others possible, and has given the hypothesis a positive Bayesian balance; that has made it useful.

In the case of consciousness, any theory must have a selection law, which says: 'certain events in the brain cause consciousness; all other events do not'. This has been a big unsolved problem for neural theories of consciousness, which need to define what kinds of neural events cause consciousness, and what kinds of neural events do not. It is known that most neural events in the brain do not cause consciousness [Merker 2007] (many lesions do not interrupt consciousness); but nobody has yet proposed a reasoned basis for a selection law to distinguish those neural events that cause conscious awareness, from the majority which do not. Proposals have been made; but being guesses, they must all count on the debit side of the Bayesian balance.

**Decoding Law**: If any theory of consciousness is to achieve a positive Bayesian balance, it needs to make predictions which can be compared with empirical evidence. These predictions cannot be only about the existence of consciousness; it is not enough for a theory to simply say 'consciousness exists', because that gives only about one bit of testable information, which is not enough to make a positive Bayesian balance. The predictions need to be about the properties of consciousness, and how those properties depend on the selected physical events. The theory needs to define how the properties of physical events (usually, the firing rates of selected neurons) are to be decoded to predict the properties of consciousness. It is not enough just to say: 'when these neurons fire, consciousness exists.' Which properties of the neurons firing lead to which properties of consciousness? In particular, neurons in the brain are known to represent space in complex, distorted ways; yet our 3-D spatial conscious experience is largely undistorted. This means that the distorted neural representations of space must be decoded, to remove the distortion. The decoding is necessarily complex – as complex as the encoding it unscrambles. All complexity in the decoding law is part of the complexity of the hypothesis, and counts on the debit side of the Bayesian balance.

A further challenge for neural theories is that the bridging law is expected to be a fundamental law of nature, and so should be independent of evolutionary time and of species. If it depends on some transient details of the current human brain, it would make consciousness into an idiosyncratic consequence of this particular evolutionary era – an unattractive feature in a theory.

For most theories of consciousness, when the three components on the hypothesis debit side are taken together (cognitive base assumptions, selection law, and decoding law), they typically add up to a debit of at least a hundred bits – possibly much more (for instance, 50 bits cognitive base; 20 bits selection law; 50 bits decoding law). You do not get much assumption for 50 bits. In typical theories of consciousness, the textual description of the hypothesis occupies much more than 50 bits – it takes kilobytes or more.

How can this large debit be balanced by a good fit to data, to attain a positive Bayesian balance?

## 3. Data to Test Theories of Consciousness

Chalmers [1996] has discussed the kinds of evidence available to test a theory of consciousness. We need to ensure we are talking about phenomenal consciousness (the 'what it is like' to be alive; which Chalmers calls conscious experience of something) rather than access consciousness [Block 1978] (which Chalmers calls awareness – the ability to act on some piece of information, or to report it verbally). Since there is a strong correspondence between access consciousness and phenomenal consciousness, we can use access consciousness as evidence of phenomenal consciousness.

It would be possible to split philosophical hairs about this; but if you consistently applied a skeptical philosophical strategy of 'deny the data', then no science would be possible. Without a certain amount of realism, consciousness (or any other phenomenon) is not an issue, and there would be no point in doing science. I assume the reader wants to do science.

There are several categories of empirical data about phenomenal consciousness:

1. Levels of consciousness – how they vary over time, depending on physical facts
2. Similarity scales of qualia (colours, sounds, smells…)
3. Shapes of things at the present moment in conscious awareness, from sense data
4. Consciousness of imagined things and remembered things
5. Basic consciousness of self
6. Higher order consciousness of self
7. Consciousness of thoughts, inner monologues and bodily feelings
8. Anomalies and illusions of consciousness
9. The existence of consciousness (the hard problem)

There may be other categories which I have not mentioned, and one could spend some time discussing each category. I only intend to make one point; that category (3) is a much richer source of testable data than all the other categories put together.

If we pause at any moment of the day and ask: 'what is my conscious experience at this moment?', the answer is 'I am aware of the things around me'. In particular, I am aware of their geometric shapes in three dimensions; and that geometric information has very high information content – probably thousands of bits of information at any moment.

For instance, you experience the edge of your desk as a straight line, with a high level of precision. Precision is information content; you could detect deviations from a straight line at the level of $10^{-2}$ ($2^{-6}$, or 6 bits) in any of 10 regions along the edge – which makes 60 bits, just from one aspect of your experience. There are many other aspects of experience in any small time interval, as your attention goes to different things.

The testable fact is that the geometry of your conscious experience matches the geometry of real space around you, in some respects with high precision. You can take a ruler, a taut string, or a set square and laboriously verify this if you wish. This testable fact is something that a theory of consciousness can predict; or may not predict it.

If a theory of consciousness can predict that the geometry of conscious awareness matches the geometry of real space to high precision, then that successful prediction adds positively to the Bayesian balance of the theory – adding some quantity of the order of 1000 bits or more (depending on what time intervals are believed to provide independent support for the theory). This may be sufficient in the Bayesian balance to offset the debits in the hypothesis.

The same cannot be said of any of the other categories in the list (1) – (9) above. Some of the categories, such as consciousness of self, or the existence of consciousness, have very small information content. Other categories (consciousness of qualia such as sounds, tastes and feelings) have middling information content – but for those categories, it is hard to define an objective comparison with external reality as the basis for testable predictions. They could reasonably be taken to be part of the same consciousness of space (external and internal) in category 3. Bodily feelings and sounds have shapes in space, albeit not precise shapes.

The result is that in order to attain a positive Bayesian balance, any theory of consciousness must make successful predictions of the geometric shape of conscious experience, category 3. Without that, there is no prospect of confirming any theory of consciousness.

This narrows the field of theories that can achieve a positive Bayesian balance.

## 4. Spatial Cognition and Consciousness

A large part of our conscious experience is experience of the three-dimensional space around us; so consciousness may be related to the processes in the brain which perceive, understand and exploit space – to 3-D spatial cognition. I summarise some of what is known about 3-D spatial cognition, irrespective of consciousness.

We use 3-D spatial cognition at every moment of the day. Even putting a finger round the handle of your coffee cup involves knowing where the handle is in 3-D space; using muscles in your arm to guide a finger towards the cup; knowing when to extend your finger and when to bend it round the handle; and then, tactile feedback that you have succeeded, and can go on to the next step of raising the cup.

Skilled movements like this are not uniquely human. Even small insects show a precise understanding of 3-D space, landing with exquisite speed and precision on the rim of a wine glass (using a brain with as few as $10^5$ neurons).

For most animals, the majority of the brain is devoted to spatial cognition – using sense data to understand the space around them and to control their muscles. In evolutionary terms, this is not surprising. Animals are living things that move, and they need to do it skillfully at every moment of the day. The first task for a brain is to control their muscles. There has been enormous and sustained selection pressure on brains to do this well, so it is no surprise that most of the animal brain is devoted to spatial cognition. 3-D spatial cognition is the primary task for a brain – coming before other functions like memory, learning and social behaviour.

In spite of the primacy of spatial cognition, it is still understood only in the broadest, most approximate terms. The problem of spatial cognition was posed in Marr's [1982] book, 'Vision' and he made initial steps to analyse it – in his '2 ½D sketch' and humanoid body modelling. Since then, there have been essentially no working neural cognitive models of how brains do 3-D spatial cognition. Why is this?

A possible reason if given in [Worden 2024a]. There are no working neural models of spatial cognition because, in engineering terms, it is not possible. Any neural representation of 3-D space would be too imprecise and too slow.

The theory of Bayesian cognition [Knill & Pouget 2002, Worden 2024c] implies that animals build internal Bayesian maximum likelihood models to guide their behaviour. In particular, they build a 3-D spatial model of their surroundings from vision and other senses. Some 3-D information can be inferred from stereopsis and touch, but this is limited to restricted spatial regions. More can be inferred from the Bayesian prior probability that most objects around the animal do not move, as the animal moves in space, using a computation of Shape/Structure from Motion (SFM).

[Worden 2024a] is a working computational model of 3-D spatial cognition in bees and bats, using Structure from Motion, at Marr's [1982] Level 2. It is not a neural implementation, but it can be used to explore the effect of neural memory errors on the computation. Short-term spatial memory is required to compute SFM. The results are clear-cut: the expected errors in neural spatial memory are too large to support the calculation of shape from motion.

In neural models of cognition, short-term memory is stored as stochastic neural firing rates. If components of 3-D positions are stored as firing rates, and if a neuron fires N times in some short time interval, the expected precision in spatial locations is of the order of one part in $\sqrt{N}$. For a small mammal, which needs to understand its surroundings and move within, say, 1/10 second, N is at most 50, giving a spatial precision of about one part in 7, at best. For an insect, the timescales are even shorter [Chittka 2022]. However, the model in [Worden 2024a] shows that relative spatial precisions of the order of 1 part in 100 are needed to compute shape from motion. This would require 100 times more neural firings – implying delays of 10 seconds or more, which are of course not observed.

So if spatial memory is held as neural firing rates, the required tradeoff between precision and speed is not possible. This is one reason why working neural models of spatial cognition do not exist. Nobody has ever built a neural implementation of 3-D spatial cognition, because it cannot be done.

How then do animals build internal models of the space around them? Empirically, they do it very well. [Worden 2020a, 2024b] proposes that spatial short-term memory is stored not as neural firing rates, but in a wave in the brain. This may seem a radical hypothesis, after so many years of neural cognitive modelling; but there is strong evidence to support it, summarized in section 6. If the wave hypothesis is correct, then it has far-reaching consequences for theories of consciousness. These will be explored after discussing neural theories of spatial consciousness.

## 5. The Projective Consciousness Model (PCM)

I know of only three neural theories of spatial consciousness – Trehub's [1989] theory, IIT [Tononi 2012] as applied to spatial consciousness [Haun & Tononi 2019], and the

Projective Consciousness Model (PCM) [Rudrauf et al 2017, 2022; Williford et al 2018]. This section discusses the PCM.

The PCM is an application of the Free Energy Principle (FEP)[Friston Kilner & Harrison 2006; Friston 2010] and Active Inference (AI) [Friston et al 2017]. The Free Energy Principle is a Bayesian approach to cognition, in which animals build Bayesian maximum likelihood models, using their sense data, to guide their actions. Active Inference says that the Bayesian internal models of reality are not just the result of passive reception of sense data, but that actions are chosen in part for their epistemic value – how they help to gather the sense data needed to build internal models of reality – as well as for their direct contribution to fitness. Actions are chosen to explore as well as to exploit. Active Inference defines how the tradeoffs between exploration and exploitation are made.

The PCM builds on this foundation. As Bayesian cognition is now well supported by much empirical evidence, and as the FEP/AI has been successfully applied in many domains of cognition, this means that the first part of the hypothesis debit side of the PCM (its neural/cognitive basis) is not based on unsupported hypotheses, and does not add much to the debit.

The PCM proposes that one of the Bayesian internal models built in the brain is a projective transform of the 3-D space around the animal. The reason given for proposing a projective transform is that an animal's internal model of space may need to be transformed for purposes of anticipation and planning – for instance, to understand what the model of space would be if the animal were to move in a certain way (perspectival imagination). So the PCM model of space is motivated by active inference – by what would be the impact of actions on an internal model.

The reason why a projective transform is proposed – out of all the possible transforms of 3-D space – is that a projective transform has useful mathematical properties. In particular, a projective transform of space preserves all straight lines. Straight lines are useful to animals for many purposes, such as predicting movement (which goes in straight lines unless perturbed) and object recognition.

The PCM then proposes that the internal projective model of 3-D space is the basis of phenomenal consciousness. Therefore it predicts that the geometric form of phenomenal consciousness is a projective transform of real external space. This prediction of the PCM agrees with the properties of conscious experience in an information-rich manner. For instance, it predicts – in agreement with our experience - that straight lines in conscious experience are straight lines in reality. This information-rich agreement of its core prediction with data is a strong positive contribution to its Bayesian balance.

The PCM makes other predictions about the geometric nature of consciousness – for instance, concerning perceptual illusions such as the moon illusion. These are interesting consequences of the PCM, but their contribution to its Bayesian balance is not as large as its core agreement with the spatial form of consciousness.

The FEP has a neural process model, which describes how the quantities used in the FEP/AI formalism (such as sense data, Bayesian probabilities, and gradients of probabilities) are represented by neural connections, synaptic strengths and firing rates; and how the computations of the FEP formalism are done in the brain, as neurally implemented additions, multiplications and subtractions. The PCM builds on the FEP neural process model, and locates the projective model of space (the Projective Geometry Engine) in the frontal cortical region. However, the neural process model of the PCM has not been implemented.

Thus the PCM gives a direct account of the spatial nature of conscious awareness. In this, it goes further to address the main experimental fact about consciousness (its spatial form, which is close to the form of external reality) than most theories of consciousness.

## 6. Difficulties for Neural Theories of Consciousness

As described above, the PCM has strong contributions on the credit side of its Bayesian balance, from its good agreement with the geometric form of conscious experience. On the debit side, the first part of its hypothesis debit (from its neural/cognitive basis) may not be large. What about the remaining parts of its assumption debit? There are the following difficulties, which the PCM shares with other neural theories. The PCM is used for illustration:

1. **The selection problem**: The PCM assumes that certain neurons in the prefrontal cortex – and no other neurons – give rise to conscious experience. There is no physical basis for this assumption; on the contrary, cortical columns appear to be physically similar and interchangeable in function across many parts of cortex – and not needed for consciousness [Merker 2007]. The singling out of just those neurons in prefrontal cortex associated with one representation of space (amongst the many representations of space in the neo-cortex) is arbitrary, and counts as a debit in the Bayesian balance.
2. **The precision/speed problem**: The neural process model of the FEP, which is adopted by the PCM, assumes that model quantities, such as the components of spatial positions, are represented by neural firing rates. But the FEP process model has not been implemented for the PCM, and the issue of neural precision and error rates has not been tested. The 3-D model of space underpinning our own consciousness is - at least in certain regions of

space – highly precise, with relative errors less than one part in 100, and responds to changes in less than a second. Any simple representation of space by stochastic neural firing rates – as in the FEP process model – cannot give this speed and precision. More complex alternative representations of space might be proposed; but their feasibility has not yet been demonstrated, they have not been found in the brain, and they would add considerably to the complexity of the neural computational model, and to the complexity of decoding needed to give conscious experience.

3. **The decoding problem**: All the known representations of space in the cortex are highly distorted representations (as, for instance, in V1 visual cortex) and are species-specific. Conscious experience has little spatial distortion, compared to real external space. So either the projective model of space in frontal cortex is unlike all other cortical representations of space, and is highly undistorted; or there is some complex decoding law which converts the physical firing of prefrontal neurons into our undistorted conscious experience. An undistorted neural model of space has not been detected in the frontal cortex, and seems to be unlikely; so a highly complex decoding law is needed, which would add considerably to the debit side of the PCM Bayesian balance. Any extra complexity of neural encoding of space, as required to overcome the speed/precision problem, would add more to the debit.

4. A further difficulty for the PCM, if the source of conscious experience is the frontal cortex, is that lesions of the frontal cortex do not destroy consciousness – as has been known since the case of Phineas Gage in the 19th century. Consciousness can even exist without a cortex [Merker 2007].

The decoding problem is illustrated as a 'hard triangle' in figure 2.

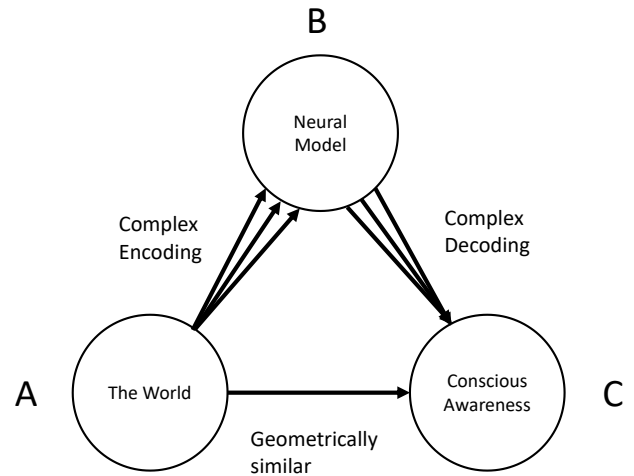

*Figure 2: The Hard Triangle. A complex encoding of spatial information from A to B implies the need for a complex decoding from B to C*

Conscious awareness (C) is very similar to real space (A). Through sense data, real space gives rise to many complex and distorted neural representations of space in the brain (B). These are complex encodings of space. So to get from (B) to (C) requires equally complex decodings.

Taking these problems into account, the PCM theory of consciousness does not yet have a positive Bayesian balance.

Other theories of spatial consciousness [Trehub 1989, Haun & Tononi 2019] have the same difficulties (1) – (3) as the PCM.

However, we might find a theory of consciousness with a positive Bayesian balance by combining ideas of the PCM with a new proposal for spatial cognition, described in the following sections.

## 7. A Wave Hypothesis of Spatial Cognition

As was described in section 4, neural theories of spatial cognition face a serious difficulty of speed and precision. Animals and people do 3-D spatial cognition very well, building a precise 3-D internal spatial model with sub-second response times. There are no known neural implementations of this, and the reasons appear to concern the speed and precision of neural spatial representations; they are simply not good enough to support a fast, precise 3-D spatial model. Neural representations of space are both too slow, and too imprecise.

[Worden 2024a] describes a working model of 3-D spatial cognition in bees and bats, which attains realistic levels of performance. This model is built at Marr's [1982] level 2 of algorithms and data structures; it is not a neural implementation. It can be used to explore what levels of errors in the representation of space are acceptable in a model of 3-D spatial cognition. The overall result[1] is that errors of the order of 1% can be tolerated; but higher levels of errors cannot. Those low error levels might be achieved by storage of spatial information not by neurons, but in a wave in the brain.

The model of [Worden 2024a] can be extended to do Bayesian multi-sensory integration, in an Aggregator model [Worden 2020b].

The wave hypothesis of spatial cognition is;

**In the brain's internal model of local 3-D space, the locations of objects are represented by a wave in the brain**.

If there is some approximately spherical volume in the brain, which can hold wave excitations, then each wave can have a different wave vector, or **k**-vector. This is a three-dimensional vector which describes both a wavelength and direction of a wave motion (**k** is orthogonal to the wave fronts). If the physics of the wave is linear, the same volume can hold many independent waves, with different **k** vectors – which do not interfere with one another.

Therefore a wave can be used to store the independent locations of many objects – objects with different locations **r**, related to the wave vectors by **k**=α**r**, where α is a constant. If we assume that:

- Each wave excitation can persist for short periods (fractions of a second)
- The minimum possible wavelength λ is small compared to the diameter D of the volume (so there is a large range of possible **k**-vectors, in all directions)
- Neurons can couple selectively to waves of different wavelength and direction, as transmitters and receivers (e.g. one neuron might have its wave receptors or transmitters aligned and spaced with the wave fronts, to be selective near one **k**-vector) then neurons could use the wave as a short-term memory for the locations of many objects. This form of spatial memory could have benefits over storage in neural firing rates:

1. The three dimensions of the wave correspond directly to the three dimensions of object positions; there is no need for any preferred direction, or for the representation to be asymmetric between directions. There is no need to choose a coordinate system. It is a direct and natural representation of positions – representing 3-vectors by 3-vectors.

2. A large number of **k**-vectors (independent object positions) - of the order of $(D/\lambda)^3$ - can be stored in the same wave volume.
3. The precision of each object location in any dimension is approximately one part in $(\lambda/D)$ – which can be smaller than one part in 100, as appears to be required for effective model building.
4. As in a hologram (which works by the same principle) there is very little spatial distortion of positions.
5. In principle, the wave can be updated or read very fast, say within a few milliseconds. There is no hard tradeoff between speed and precision.

These benefits overcome the problems of speed and precision from neural storage.

There are other possible benefits, if we make further assumptions about the wave and how neurons couple to it. Further potential benefits are:

6. A single neuron's coupling to the wave might be tunable to different wave vectors – if the sensitivities of its wave transducers (possibly in its synapses) can be altered, or can be given phase delays, by a separate steering signal. This gives a way for the wave to be used for selective **spatial steering** of signals – something which is needed for dynamic routing of information, for instance from sense organs to specialized pattern recognition modules [Treisman & Gelade 1980; Treisman 1998; Worden 2020b]
7. Individual neurons could be tuned not just to specific wave vectors (specific represented positions) but also partially de-tuned to represent the uncertainty in object positions, as a Gaussian-like spread of wave vectors.
8. If the phases of the waves could be controlled, physical addition of wave amplitudes from different sources might directly represent the addition of negative log likelihoods (or free energies [Friston 2010]). This is required to find Bayesian maximum likelihoods, and in the aggregator model [Worden 2020b], for multi-sensory integration.
9. If the wave has several internal degrees of freedom (as, for instance, polarization of an electromagnetic wave gives two degrees of freedom), different degrees of freedom could represent different attributes of objects, such as colour or hardness.

If spatial positions are stored in a wave in the brain, there must be some minimum possible wavelength $\lambda_{min}$ that neurons can couple to; which implies that there is a maximum **k**-vector, and a maximum distance that can be represented. This is a problem for representing very large distances, which animals sometimes need to do. So the wave excitation cannot represent Euclidean space directly, but probably represents a transform of Euclidean space, which has evolved to minimise geometric distortions. In this respect, projective transforms of space [Rudrauf et al 2017, 2022] are particularly appropriate, as they can reduce infinite distances to finite distances, and they preserve straight lines; straight lines are important for controlling motion and recognizing shapes. So the wave storage may use some near-projective transform of Euclidean space.

---

[1] This result depends on the parameters used in the model, such as the assumed visual acuity. The program to run the model can be downloaded in www.bayeslanguage.org/bb/BB.zip ., and the parameters can be varied.

A purely projective transform puts all points at infinite distance in real space onto a plane in projected space. In a Fourier transform (wave representation or projected space) a distant plane has a maximum value of one component of the wave vector **k**. However, a minimum wavelength (which we might expect in the brain) is a maximum modulus of the wave vector **k**, irrespective of its direction. So rather than a pure projective transform, the brain might use a near-projective transform, in which objects at very large distances appear to be on the surface of a sphere, rather than in a plane. This near-projective transform would preserve straight lines approximately, but not exactly.

This may be why we see the stars as a spherical canopy in the sky, and why perception has minor distortions of Euclidean geometry [Koenderlink et al 2000]

The wave representation of space works best in an (intermittently) allocentric frame of reference, where any stationary object is represented by an unchanging wave vector. This simplifies the storage of non-moving things, which are in the majority.

Storage in a wave has potential for a spatial short-term memory, which could be greatly superior to neural memory. Many details remain to be worked out – including:

a) What is the physical nature of the wave?
b) How can the wave be sustained for the required (sub-second) times?
c) What is the source of energy for the wave?
d) How do neurons couple to the wave? What genes and proteins are involved?
e) Can neurons have steerable coupling to the wave?
f) Where in the brain does the wave reside?

There are candidate answers for (f), described in the next section.

## 8. Evidence for Wave Storage of Spatial Information in Brains

There is strong indirect evidence for a wave excitation in the brain. [Worden 2020, 2024b] proposed that:

- There may be a wave excitation in the mammalian thalamus, storing spatial information
- The central body of the insect brain may play the same role

Supporting evidence is described in the papers. Here I summarise the most important evidence, common to the mammalian thalamus and the insect central body:

1. Both the thalamus and the central body have a simple, near-spherical anatomical form, which is remarkably well conserved across a wide range of species [Sherman & Guillery 2006; Jones 2007; Heinze et al. 2023]. A near-spherical shape is well suited to hold a wave in three dimensions. This shape is in marked contrast to the contorted, variable and species-specific shapes of most parts of the brain (such as the neo-cortex, the hippocampus, or the mushroom bodies). This strongly suggests that in the thalamus and the insect central body (and not in other parts of the brains), the shape serves some purpose other than neural synaptic connections is going on. That purpose could be to hold a wave – because the shape is right for it.
2. Both the thalamus and the central body are centrally located, and richly connected to other parts of the brain – suggesting that they both act in an integrating hub role for some important type of cognition. Spatial cognition with multi-sensory integration is such a function, because it underpins everything an animal does in its life, and is a key determinant of fitness.
3. Both the thalamus and the central body are innervated by sense data of every modality – except, remarkably, olfaction[2]. Smell is of little use for the precise, fast, location of things in a 3-D model of space. This may explain why smell is not strongly linked to either of those central brain parts.
4. Both the thalamus and the central body are closely linked to consciousness [Baars 1988, Chittka 2022]. Lesions in the thalamus, or targeted predator injection attacks on the insect central body, cause cessation of consciousness.

A typical shape of the insect central body is shown in figure 3. This shows the outer fan-shaped body and the inner elliptical body.

---

[2] Olfaction, unlike all other sense data, does not pass through the thalamus on its way to the cortex. The insect central body has no direct connections to the mushroom bodies, centres of olfactory learning. However, it may have other connections to olfaction.

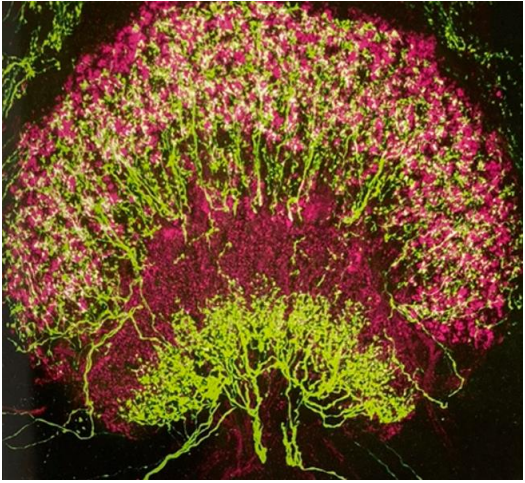

*Figure 1: Central Body in the brain of a honey bee, from [Strausfeld 2012]*

Some typical shapes of mammalian thalami are shown in figure 4.

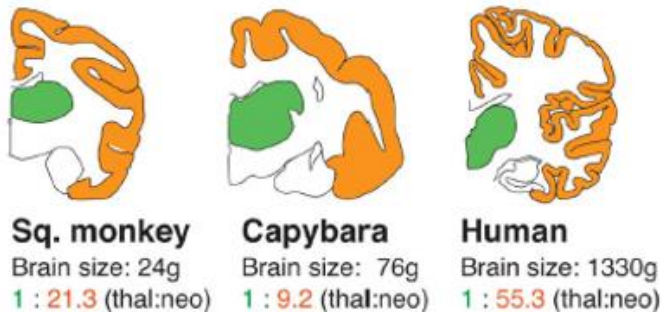

*Figure 4: Shape and size of the thalamus in various species, from [Hailey & Krubitzer 2019]*

It can be seen that the shapes of both the central body and the thalamus are nearly spherical, which is very distinctive (unusual in the brain) and is well suited to hold a wave. Both shapes are remarkably conserved across many species.

In the mammalian thalamus, there is further strong evidence for a wave [Worden 2010]. Without a wave, the anatomy of the thalamus does not make sense. The thalamus is a cluster of centrally-placed thalamic nuclei, with few or no connections between nuclei [Sherman & Guillery 2006, Jones 2007]. In terms of brain energy consumption and neural connections, this does not make sense. The same neural connectivity (i.e. the same neural computation) could be obtained with less brain energy consumption (less net axon length) if the thalamic nuclei were separated, migrating outwards towards the cortex. The anatomy of the thalamus only makes sense if the thalamic nuclei need to stay close together. They would need to stay close, to be immersed in the same wave.

There is similar supportive evidence from the distinctive shell-like form of the Thalamic Reticular Nucleus (TRN), wrapped around the dorsal thalamus [Worden 2014].

Taken together, this evidence is significant. The wave hypothesis accounts for striking neuro-anatomical facts, which are not explained in a purely neural model of the brain.

Locating the wave in the mammalian thalamus is consistent with proposals that the thalamus acts as a blackboard for the integration of many cognitive functions [Baars 1988; Mumford 1991; Worden, Bennett & Neascu 2022]. It is also consistent with the central location of the thalamus, with its prominent connections to many kinds of sense data and many regions of cortex.

The wave hypothesis of spatial cognition has implications for theories of consciousness, described in the next section.

## 9. A Projective Wave Theory of Consciousness

This theory of consciousness, based on the PCM, proposes that:

**Consciousness of space arises from a wave in the brain, which holds a near-projective 3-D model of local space.**

To describe the theory, I consider the Bayesian balance between its hypothesis complexity and its fit to the data.

The hypothesis complexity has three components:

1. Neural/cognitive basis
2. Selection rule
3. Decoding rule

**The neural/cognitive basis** of the theory is a theory of spatial cognition which, while novel, accounts for animals' very capable spatial cognition in a way that purely neural theories currently do not. It does so at the cost of proposing a wave excitation in the brain, which has not yet been observed. However, there are good reasons why the wave may not yet have been observed (it is expected to have very low intensity, and we have not known how to look for it), and there is persuasive indirect evidence that a wave exists, in the central body of the insect brain and in the mammalian thalamus. So while the hypothesis debit of the neural cognitive basis of the wave theory of consciousness is not zero, it is not prohibitively large.

The **Selection Rule** of the theory is just that the wave in the brain is the origin of consciousness; and that neural activity in the brain has nothing to do with consciousness. This is much simpler than the selection rule of any neural theory; it is also consistent with a single, universal, timeless and species-independent bridging law between physical events and consciousness [Chalmers 1996]. In these respects, the theory is more satisfactory than neural theories, and the debit of its Bayesian balance is small.

The **Decoding Rule** of the theory uses the mathematics of the Fourier transform. The amplitude of the wave in the brain is a Fourier transform of projected 3-D local space; the spatial form of conscious experience is just the inverse Fourier transform of the wave[3]. This is a very simple decoding law, comparable to the applications of the Fourier transform in many fields of science (notably in quantum mechanics); so its contribution to the Bayesian debit of the theory is small. The decoding law is consistent with a single, universal, timeless and species-independent bridging law between physical events and consciousness.

So in terms of hypothesis complexity, the Projective Wave Theory of Consciousness only has a modest debit. How does it fare on the credit side, in its fit to the data?

The central prediction of the theory is that the spatial form of consciousness closely resembles the geometry of real 3-D space – straight lines in conscious experience are straight lines in real space, and so on. This prediction agrees very well with the data from our conscious experience, making a large positive contribution (possibly thousands of bits) to its Bayesian balance.

A further prediction of the theory, agreeing with the data, is that lesions of the thalamus destroy consciousness [Baars 1988, Jones 2007, Sherman & Guillery 2006] – unlike lesions in almost all other parts of the brain.

In sum, the Bayesian balance of the Wave theory of consciousness is positive – unlike all neural theories of consciousness. This is reason to believe it is more than just a hypothesis.

As a scientific theory, the Projective Wave Theory has another major benefit over current neural theories of consciousness – that it is falsifiable. Any theory should be falsifiable [Popper 1934]. If thorough investigation shows that spatial cognition is not done using a wave in the brain, then the Projective Wave Theory will be wrong. It can be falsified by evidence not pertaining to consciousness.

In an analogy, Pauli [1930] predicted the existence of the neutrino, at a time when he thought neutrinos could never be detected. He might have been wrong, but neutrinos were directly detected more than twenty years later [Cowan Reines et al. 1956]. Similarly, dark matter and dark energy are now being intensively investigated – both by people who would like to show they exist, and in experiments whose aim to show they do not.

## 10. Physical Nature of the Wave in the Brain

The Projective Wave theory of consciousness depends on the existence of a wave in the brain. To detect a wave, we need to consider its physical nature.

It is tempting to propose that the wave is something we now understand – specifically, that it is an electromagnetic wave [e.g. Pockett 2000, McFadden 2002]. I suggest that this is unlikely, for two main reasons.

A. The biological purpose of the wave is to store spatial information for short periods of time – say, up to a second – using as little metabolic energy as possible. To do this, it will have evolved to the lowest possible intensity, while storing signals above the level of background noise. Neuronal firing creates quasi-random electromagnetic background noise in the brain, so an electromagnetic wave cannot have evolved to have low intensity. An electromagnetic wave would require high intensity to overcome background noise in the brain.

B. Our conscious experience includes qualia such as colour at any location in our experienced space. If consciousness arises from a wave, then the wave must have many internal degrees of freedom, to represent the qualia at any location in experienced space. An electromagnetic wave only has two degrees of freedom (in its polarisation), which is not sufficient to account for the many qualia we experience.

So, like a drunkard searching for his keys, looking for an electromagnetic wave in the brain might be to look under the wrong lamp-post. We need to look for something less familiar.

In this respect, we can be encouraged by two developments: (a) the many different quantised excitations that have been discovered in solid-state physics; and (b) recent evidence for the existence of coherent quantum effects in biological matter and the brain [Kerskens & Perez 2022].

The purpose of the wave is to store information for times which, on the timescales of molecular interactions and thermal fluctuations, are very long. So it may depend on phenomena that are well insulated from the dominant chemical and electrical activity in the brain – so that their quantum coherence times can be much longer than the timeframes of the order of $10^{-20}$ second which are implied for neuron states by quantum decoherence arguments [Tegmark 1999, Hepp & Koch 2007; Schlosshauer 2010].

---

[3] Two successive Fourier transforms of a spatial distribution result in the same spatial distribution. So the second transform is called the inverse transform.

For instance, states of nuclear spin could have a high level of insulation from cellular events, enabling them to remain coherent for longer times. How neurons might couple to nuclear spins is at present unknown; but there are intriguing hints of the relevance of spin states to neural activity, in that the effectiveness of Xenon as an anaesthetic depends on the nuclear spin of different isotopes of Xenon [Li et al 2018].

If the evolution of wave-based spatial memory put it on a trajectory of increasing quantum coherence times, the end point of this trajectory could be a Bose-Einstein Condensate (BEC) [Bose 1924; Einstein 2025] BECs such as superfluids can store information (as rotons) indefinitely [Feynman, Layton & Sands 1965; Pitaevskii & Stringari 2010]. Identifying the wave as an excitation of a BEC would have several advantages:

- A BEC is not immune to decoherence effects (long-lived 'Schrodinger Cat' BEC states are not observed). But some information in a BEC state is immune to decoherence and thermal noise. This is experimentally well established and theoretically understood. So the decoherence arguments that quantum effects are not important in the brain [Tegmark 1999, Hepp & Koch 2006] do not apply to a BEC.
- Because BECs can store information for long times, they are an ideal substrate for memory. They could have been used for memory for very long evolutionary times – even in single-celled animals.
- There are theoretical models of biological BECs at high temperature [Frohlich 1968], sustained by molecular pumping to produce a population inversion, analogous to coherence in a laser.

BECs are a fundamental and distinctive state of matter, and they are very rare in nature. If BECs are the substrate of consciousness [Marshall 1989; Worden 1999], the selection rule for consciousness has a very simple form: Consciousness exists in BECs, and nowhere else. Any BEC is conscious of the information in itself. This simple bridging rule for consciousness would be intellectually satisfying, and would avoid the taint of pan-psychism. It would also limit our expectations of sentient AI.

## 11. The Evolution of Consciousness

Many theories of consciousness face a fundamental difficulty, that they give no account of how or why consciousness evolved. The theories imply that consciousness (unlike all other properties of life) was not designed by evolution; that the form of consciousness is an accident.

The difficulty arises from the neural causal model of the brain. If neurons in the brain act as a closed, classical, local computing system [Tegmark 1999, Hepp & Koch 2006], the results of those computations drive natural selection – so neurons and brains have been designed by evolution. Neurons also cause consciousness, but consciousness is outside the closed classical neural model of the brain. So consciousness has no causal effect on neural computation, and therefore has no effects on survival. There are then no selection pressures to evolve the form of consciousness.

This appears to be wrong, because consciousness is faithful to the real world. Has the fidelity of consciousness not been designed by evolution? Does it serve no purpose? Is the form of consciousness just an accident? That seems to be an outrageous coincidence – in other words, it has a highly negative Bayesian balance; and is not to be believed.

The projective wave theory avoids this difficulty, and gives sound evolutionary reasons why consciousness is the way it is.

Spatial cognition is the most important function of a brain. The wave enables brains to build an internal 3-D model of local space, to control movement every moment of the day. This model is essential to survival, and has been under intense selection pressure for more than 500 million years, since the Cambrian era – so the wave model of space is exquisitely designed to help animals survive. Consciousness results from the wave by a Fourier transform – so the wave and consciousness have been exquisitely designed by evolution, even though consciousness itself has no causal effect on survival. The wave does have effects. The form of consciousness is then not an accident.

The theory also implies that consciousness is as old as spatial cognition – at least 500 million years old. It exists in all animal phyla, from insects to mammals.

## 12. The Unity of Consciousness

The projective wave theory of consciousness gives a simple answer to a question that is a difficulty for neural theories of consciousness – why do we have only one consciousness, and not many parallel consciousnesses? In the wave theory, we have only one consciousness because there is only one wave in the brain.

If neurons were the origin of consciousness, it would be perfectly feasible for one person to have many consciousnesses at the same time – there are enough neurons to do it, and brains are known to do many things in parallel[4]. Having several consciousnesses would undoubtedly have evolutionary benefits – when facing some immediate danger, we could consciously imagine and plan several different possible physical responses in parallel, and choose the best one. We do not do that, which strongly suggests that neurons are not the source of consciousness; but that consciousness depends on some bottleneck in the brain, that cannot be replicated. That bottleneck is the wave in the brain, which is the only source of consciousness.

---

[4] For instance, in understanding any sentence you hear, you rapidly and in parraller retrieve many possible wods to fit the sounds, and you try out in parallel different parses of the same sentence, finding the most likely meaning.

In this theory, the most important function of a brain is to plan and carry out physical actions. That requires a precise representation of 3-D physical space, and neurons cannot do that. The most precious resource in the brain is the single wave excitation which does. A single wave in the brain gives a single consciousness.

## 13. Consciousness is a Scientific Problem

The projective wave theory depends on the existence of a wave in the brain, representing 3-D space. If this wave does not exist, or is shown not to be necessary, then the theory is wrong. On the other hand, if the wave exists, then there are compelling reasons to link it to consciousness. There is such a good match between the properties required of the wave, and the properties of our spatial conscious experience, that it would be perverse not to link the two.

The projective wave theory of consciousness stands or falls by the wave. The existence of a wave in the brain is a question of neuroscience, not of philosophy. There is a good chance that it could be answered with some confidence from neuroscientific evidence within, say, the next ten years. This would be a refreshing change, after hundreds of years of philosophical debate about consciousness.

Philosophers may object to this impending resolution, saying: "What about the Hard Problem of Consciousness? It will not have gone away." Indeed it will not. But there are hard problems at the base of all science. There is a hard problem of physics – why does matter exist? There is a hard problem of cosmology – why do universes exist? These questions will never be answered, because there is an infinite regress of 'why' questions. In the case of physics, we have found a series of answers so far: matter => molecules => atoms => protons => quarks; but we can only extend the series as far as there is experimental data, to give our hypotheses a positive Bayesian balance. It will be the same with consciousness.

## 14. Conclusions

For many years, theoretical approaches to consciousness have divided into two schools, which can be characterized as 'all you need is neurons' versus 'something else is needed'. The debate between the two schools is unresolved, largely because there has been no credible theory of consciousness in either school.

Any theory stands or falls by its agreement with experimental data, and theories of consciousness have suffered from a dearth of data to test them. This paper proposes that the main source of data to test any theory of consciousness is the spatial, geometric form of conscious experience. Very few theories attempt to predict this form. One which does so is the Projective Consciousness Model (PCM) – which predicts that consciousness is a projective transform of real 3-D space, in good agreement with our conscious experience.

While the PCM succeeds in this respect, it has other difficulties, arising from its neural basis. This paper has described those difficulties – which arise in any neural theory of consciousness – and has proposed a way to solve them. The difficulties can be avoided if the brain's internal Bayesian model of 3-D space is stored in a wave excitation; and if the wave is the source of phenomenal consciousness.

The resulting Projective Wave theory of consciousness has a positive Bayesian balance, between the complexity of its assumptions and the data it agrees with. This means that – unlike many theories of consciousness - there is a significant chance that it is correct. In this respect it at least deserves to be investigated further – even for the legitimate scientific aim of proving it wrong.

To explore the theory further, there are many possible lines of investigation, such as:

- **Theoretical models of spatial cognition**: Is there a need for some non-neural storage of spatial information, such as a wave, or could a purely neural model of spatial cognition do the job?
- **Neuro-physiological evidence for a wave**: Further investigating the mammalian thalamus and the central body of the insect brain (for instance using connectome data), do the current indications of wave storage persist, or is there some more conventional explanation?
- **Bio-physical substrate for a wave**: what bio-physical mechanisms could support a wave, with sufficient insulation from brain activity to store information for fractions of a second, but with coupling to neurons? Might the wave be supported on a BEC?
- **Genetics and proteomics**: Is there some distinctive and localised genetic or protein signature (e.g. in the thalamus or the insect central body) of neurons coupling to a wave?

Whether or not the wave hypothesis is confirmed, these investigations could lead to important advances - not just in the understanding of consciousness, but also in the neurophysiology of spatial cognition.

If the wave hypothesis of cognition and consciousness was confirmed, it would usher in a new era in our understanding of the brain. If it is wrong, it should not take long to show that it is wrong. It has the virtue of being falsifiable.